\documentclass[aps,pra,amsmath,amssymb,preprint,superscriptaddress,showpacs]{revtex4-1}
\usepackage{mathrsfs}
\usepackage{graphicx}

\newcommand{\Ev}[0]{\mathbf{E}}
\newcommand{\Bv}[0]{\mathbf{B}}
\newcommand{\Dv}[0]{\mathbf{D}}
\newcommand{\Hv}[0]{\mathbf{H}}
\newcommand{\Pv}[0]{\mathbf{P}}
\newcommand{\Mv}[0]{\mathbf{M}}
\newcommand{\Sv}[0]{\mathbf{S}}
\newcommand{\Eop}[0]{\hat{\mathbf{E}}}
\newcommand{\Bop}[0]{\hat{\mathbf{B}}}
\newcommand{\Dop}[0]{\hat{\mathbf{D}}}
\newcommand{\Hop}[0]{\hat{\mathbf{H}}}
\newcommand{\Pop}[0]{\hat{\mathbf{P}}}
\newcommand{\Mop}[0]{\hat{\mathbf{M}}}
\newcommand{\rv}[0]{\mathbf{r}}
\newcommand{\dV}[0]{{\rm d}V}
\newcommand{\dt}[0]{{\rm d}t}
\newcommand{\hbsr}[0]{\hbar^{\frac{1}{2}}}
\newcommand{\hbsri}[0]{\hbar^{-\frac{1}{2}}}
\newcommand{\curl}[0]{\nabla\times}
\newcommand{\dive}[0]{\nabla\cdot}

\newcommand{\intV}[0]{\int{\rm d}V}
\newcommand{\inth}[0]{\int^{\infty}_0}
\newcommand{\intf}[0]{\int^{\infty}_{-\infty}}
\newcommand{\ham}[0]{\hat{\rm H}}
\newcommand{\om}[0]{\Omega}
\newcommand{\dom}[0]{\frac{{\rm d}\Omega}{2\pi}}
\newcommand{\si}[0]{\sigma}
\newcommand{\la}[0]{\lambda}
\newcommand{\aop}[0]{\hat{a}}
\newcommand{\aopa}[0]{\hat{a}^{\dagger}}
\newcommand{\bop}[0]{\hat{\mathbf{b}}}
\newcommand{\bopa}[0]{\hat{\mathbf{b}}^{\dagger}}

\newcommand{\cop}[0]{\hat{c}}
\newcommand{\copa}[0]{\hat{c}^{\dagger}}
\newcommand{\sop}[0]{\hat{s}}
\newcommand{\sopa}[0]{\hat{s}^{\dagger}}
\newcommand{\Lv}[0]{\boldsymbol{\Lambda}}
\newcommand{\Lvb}[0]{\bar{\boldsymbol{\Lambda}}}
\newcommand{\Lve}[0]{\boldsymbol{\Lambda}^{\rm e}}
\newcommand{\Lvm}[0]{\boldsymbol{\Lambda}^{\rm m}}
\newcommand{\Lveb}[0]{\bar{\boldsymbol{\Lambda}}^{\rm e}}
\newcommand{\Lvmb}[0]{\bar{\boldsymbol{\Lambda}}^{\rm m}}
\newcommand{\Gvee}[0]{\boldsymbol{\Gamma}^{\rm ee}}
\newcommand{\Gvem}[0]{\boldsymbol{\Gamma}^{\rm em}}
\newcommand{\Gvme}[0]{\boldsymbol{\Gamma}^{\rm me}}
\newcommand{\Gvmm}[0]{\boldsymbol{\Gamma}^{\rm mm}}
\newcommand{\Gv}[0]{\boldsymbol{\Gamma}}
\newcommand{\Gvb}[0]{\bar{\boldsymbol{\Gamma}}}
\newcommand{\Gvh}[0]{\boldsymbol{\Gamma}_{\rm disp}}
\newcommand{\Gva}[0]{\boldsymbol{\Gamma}_{\rm diss}}
\newcommand{\Gvbh}[0]{\bar{\boldsymbol{\Gamma}}_{\rm disp}}
\newcommand{\Gvba}[0]{\bar{\boldsymbol{\Gamma}}_{\rm diss}}

\newcommand{\lv}[0]{\boldsymbol{\lambda}}
\newcommand{\av}[0]{\boldsymbol{\alpha}}
\newcommand{\bv}[0]{\boldsymbol{\beta}}

\newcommand{\rhv}[0]{\boldsymbol{\rho}}

\newcommand{\Hvt}[0]{\mathbf{H}^{\rm TP}}
\newcommand{\Evt}[0]{\mathbf{E}^{\rm TP}}
\newcommand{\Hvtc}[0]{\mathbf{H}^{\rm TP*}}
\newcommand{\Evtc}[0]{\mathbf{E}^{\rm TP*}}

\newcommand{\curlv}[0]{\boldsymbol{\nabla}\times}

\newcommand{\Uv}[0]{\mathbf{U}}
\newcommand{\Uvb}[0]{\bar{\mathbf{U}}}

\begin{document}

\title{Canonical quantization of macroscopic electrodynamics in a linear, inhomogeneous magneto-electric medium}

\author{A. C. Judge}
\email[Corresponding author: ]{a.judge@physics.usyd.edu.au}
\affiliation{Centre for Ultrahigh bandwidth Devices for Optical Systems (CUDOS)}
\affiliation{Institute of Photonics and Optical Science (IPOS), School of Physics, The University of Sydney, NSW 2006, Australia}

\author{M. J. Steel}
\affiliation{Centre for Ultrahigh bandwidth Devices for Optical Systems (CUDOS)}
\affiliation{MQ Photonics Research Centre, Department of Physics and Astronomy, Macquarie University, NSW 2109, Australia}

\author{J. E. Sipe}
\affiliation{Department of Physics and Institute for Optical Sciences, University of Toronto, Toronto, Ontario, Canada M5S 1A7}

\author{C. M. de Sterke}
\affiliation{Centre for Ultrahigh bandwidth Devices for Optical Systems (CUDOS)}
\affiliation{Institute of Photonics and Optical Science (IPOS), School of Physics, The University of Sydney, NSW 2006, Australia}

\date{\today}

\begin{abstract}
We present a canonical quantization of macroscopic electrodynamics. The results apply to inhomogeneous media with a broad class of linear magneto-electric responses which are consistent with the Kramers-Kronig and Onsager relations. Through its ability to accommodate strong dispersion and loss, our theory provides a rigorous foundation for the study of quantum optical processes in structures incorporating metamaterials, provided these may be modeled as magneto-electric media. Previous canonical treatments of dielectric and magneto-dielectric media have expressed the electromagnetic field operators in either a Green function or mode expansion representation. Here we present our results in the mode expansion picture with a view to applications in guided wave and cavity quantum optics.
\end{abstract}


\maketitle

\section{Introduction\label{intro}}
The power of the classical theory of the electrodynamics of continuous media depends on capturing the detailed properties of the medium by a small number of spatially-averaged fields and effective response functions, such as the linear and nonlinear electric susceptibility. The response functions may take a variety of forms depending on the type of system under study (e.g., dielectric, magnetic, magneto-electric, optically active, etc.) but in all cases we replace the microscopic interactions of an enormous number of charges by a few effective macroscopic functions satisfying some general restrictions including the Kramers-Kronig and Onsager relations. In this way, many-body systems that would be impossibly difficult to analyze directly become easily tractable.

The use of effective fields and response functions is just as helpful in many types of many-body quantum theories and the concept of a quantized theory of macroscopic electrodynamics has held appeal for many authors. However, handling effective response functions in quantum mechanics can be challenging because, at least in a unitary evolution picture governed by a Hamiltonian, a quantum treatment involving the electromagnetic field is incompatible with dissipation, while dissipation is typically one of the key effects in complex systems. In the electrodynamics of macroscopic media, the relation between dispersion and dissipation, or loss, through the Kramers-Kronig relations is of central importance, and a fully quantum theory must account for it correctly. Moreover, since many important materials show very strong dispersion, approximate treatments of dispersion can have only limited validity.

Consequently, although quantization of the vacuum field was achieved soon after the formulation of quantum mechanics and the corresponding treatment for electromagnetic materials was considered soon afterwards, complete formulations of such a theory have only emerged in the last decade or so. These theories are timely since metamaterials (MM's) with unusual dispersion and significant loss are becoming increasingly common and useful. The recent studies of spontaneous emission and other phenomena in hyperbolic media \cite{Poddubny:2011}, for instance, suggest that quantum descriptions of MM's and negative media will grow rapidly in importance.

A satisfactory quantization of electrodynamics should have the following properties: it should be consistent with the Kramers-Kronig and Onsager restrictions on the response functions; recover the Maxwell equations in the classical limit; preserve the correct commutation relations between the electromagnetic field operators; and as far as possible accommodate otherwise arbitrary constitutive relations. A canonical quantization of the classical theory with given constitutive relations meets these requirements. Specifically, one must identify a Hamiltonian equal to the energy of the system and expressed in terms of the conjugate variables which, when combined with the commutation relations, yields the quantum analogue of the classical equations; in this case, the Maxwell equations. Efforts to perform such a quantization chart a long history commencing with the work of Jauch and Watson on a covariant quantum theory of linear, homogeneous, nondispersive dielectrics \cite{Jauch:1948}. Towards the same end, Drummond presented a canonical treatment of a nonlinear, dispersive, but nonabsorbing dielectric by assuming that the linear susceptibility may be approximated by a truncated Taylor expansion over a narrow bandwidth \cite{Drummond:1990}. In attempts to treat causal, absorptive media, the main challenge has been reconciling the temporal nonlocality inherent to a causal theory of electrodynamics with what must be a temporally local Hamiltonian formalism. This task was completed for a linear, homogeneous, absorbing dielectric by Huttner and Barnett \cite{Huttner:1992} who added additional degrees of freedom to the system. In their model, following the tradition of Hopfield \cite{Hopfield:1958}, the electromagnetic field is coupled to a uniform spatial distribution of simple harmonic oscillators. In addition, Huttner and Barnett introduced a reservoir of oscillators, coupled to the medium oscillators, to facilitate dissipation. However, their representation of the canonical variables using spatial Fourier transforms leads to a cumbersome theory when applied to inhomogeneous media. Furthermore, the two tiered system of oscillators has subsequently been found to be unnecessary \cite{Bhat:2006}.

An alternative phenomenological approach to quantization focuses on preserving the commutation relations of the electromagnetic field operators. This is accomplished by invoking the fluctuation-dissipation theorem to introduce source terms into the Maxwell equations corresponding to quantum noise currents \cite{Matloob:1995,Gruner:1996,Dung:1998}. In transferring to an operator formalism, these noise currents are associated with a set of bosonic fields which ensure that the commutation relations for the electromagnetic field variables are satisfied. These source terms then lead to a Green function representation for the electromagnetic field operators. Several of the canonical treatments of macroscopic quantum electrodynamics have been motivated, at least in part, by a desire to validate this phenomenological approach. Towards this purpose, Suttorp and Wubs \cite{Suttorp:2004} presented a canonical quantization scheme rooted in the Huttner and Barnett model, but extended to the case of an inhomogeneous, absorptive dielectric. It was then shown by Bhat and Sipe \cite{Bhat:2006} that the Huttner and Barnett approach could be refined by discarding the medium oscillators and coupling the reservoir directly to the electromagnetic field. In addition, \emph{magneto-dielectric} media have been considered, where the response is described by an electric permittivity $\epsilon$ and a magnetic permeability $\mu$. Specifically, a canonical treament \cite{Philbin:2010} has made contact with the application of the noise current formalism to a magneto-dielectric \cite{Dung:2003}, while spatially dispersive dielectrics likewise have been approached canonically \cite{Suttorp:2007}.

Once a quantum theory of macroscopic electrodynamics has been established via the canonical route, the door is then opened to the rigorous treatment of quantum electrodynamical processes involving dispersive and lossy bulk media. An example of such a process is the Casimir-Lifshitz effect \cite{Casimir:1948,Landau:1980b} whereby forces on solid bodies arise as a result of intrinsically quantum mechanical fluctuations in the electromagnetic field. Although the best known prediction of this theory is an attractive force between two parallel conducting plates \cite{Casimir:1948}, the possibility of repulsive Casimir forces \cite{Rosa:2008} has arisen with the consideration of left-handed media (LHM). Furthermore, with regard to the study of spontaneous processes, LHM offer novel opportunities in the tailoring of spontaneous emission by atoms \cite{Yao:2009,Noginov:2010} as well as phase matching in nonlinear optical processes \cite{Shadrivov:2006,Popov:2009,Elyutin:2010}. In the absence of such materials in nature, LHM are realised through artificial MM's consisting of structures engineered, for optical wavelengths, on the nanoscale \cite{Shalaev:2007}. In order to treat these materials as bulk constituents in an optical system, some process of homogenization must be performed whereby the electromagnetic response of the sub-wavelength structure is expressed by effective parameters for an equivalent continuous medium. Standard parameter retrieval techniques, however, frequently return results which appear to violate basic considerations such as causality and energy conservation \cite{Koschny:2003,Smith:2005}, initiating a debate on the thermodynamic validity and physical interpretation of the effective constitutive parameters \cite{Markel:2008}. It is argued \cite{Alu:2011} that the problem lies in a neglect of spatial dispersion which may be remedied to some extent by modeling the MM as a \emph{magneto-electric} medium where the polarization and magnetization each depend upon both the electric and magnetic fields. Thus, a magneto-electric response may be a general property of any plausible MM realisation of a LHM. Furthermore, the unusual interaction of light with left-handed MM's arises from their resonant properties \cite{Shalaev:2007}, which implies the presence of strong dispersion in the frequency ranges of interest for any application which exploits attributes unique to these media. Strong dispersion and, through the Kramers-Kronig relations, strong absorption, is therefore inherent to MM based realizations of LHM. A quantum treatment of the electromagnetic field in such materials must therefore include causal, magneto-electric constitutive relations representing a complete description of dispersion and loss, as opposed to a perturbative approach.

To our knowledge, a canonical quantization of electrodynamics in a magneto-electric medium has not been presented. To achieve this, we must identify a Hamiltonian operator which is consistent with macroscopic electrodynamics and yields the desired causal constitutive relations. As a first step towards such a goal, the oscillator model employed in the magneto-dielectric case by Philbin \cite{Philbin:2010} was generalized in a Lagrangian picture to encompass a magneto-electric medium by Horsley \cite{Horsley:2011}. However, no construction of the corresponding Hamiltonian was attempted.

In this paper we present a Hamiltonian operator which, with the standard commutation relations, is consistent with macroscopic electrodynamics in a causal, linear, inhomogeneous, magneto-electric medium. In order to allow for dissipation, degrees of freedom corresponding to the medium are introduced as bosonic excitations which are then coupled to the electromagnetic field variables in a bilinear fashion. This generalization of the interaction Hamiltonian constructed previously for a dielectric medium \cite{Bhat:2006} allows for the treatment of magneto-electric responses. It is possible to define this coupling to permit a description of materials where absorption is absent below a cut-off frequency $\om_c$, such as below the band-gap in semi-conductors \cite{Bhat:2006}. In the interests of simplicity we do not make this provision here, and rather assume the presence of absorption at all frequencies. Our aim is to obtain the dressed eigen-operators of the system, the polariton operators, from which the electromagnetic field operators may then be constructed. A significant simplification in the dynamics of the system is thus obtained due to the harmonic time dependence of the polariton operators. In pursuing this aim we introduce modal polariton operators (i.e. polariton operators independent of field point $\rv$) and derive corresponding mode field distributions, thus separating the spatial dependence of the fields from the time dependent polariton operators. This allows us to express the electromagnetic field operators in the form of a modal expansion where the polariton operators appear in the place of mode amplitudes. The results thus obtained constitute a complete description of the electromagnetic field operators in a broad class of a linear, causal magneto-electric media in the absence of a band-gap.

This paper is structured as follows. In Section \ref{classicalsection} we outline the classical theory of macroscopic electrodynamics to which the quantum theory must correspond, with particular attention given to energy transfer and the general properties of the susceptibility tensors describing the response of a causal magneto-electric medium. In Section \ref{hamiltoniansection} we introduce our Hamiltonian operator consisting of the electromagnetic field coupled to the bosonic excitations of a model medium. We then show in Section \ref{qmesection} how this Hamitonian leads to the quantum analogue of the classical theory with constitutive relations expressed in terms of susceptibilities which possess the Kramers-Kronig and Onsager properties required of their classical counterparts. In Section \ref{polaritonsection} we introduce the eigen-operators of the collective Hamiltonian and use them to construct solutions for the electromagnetic field operators. Finally, a general discussion is expounded in Section \ref{discussionsection}.

\section{\label{quantsection}Quantization}
\subsection{\label{classicalsection}Classical macroscopic electrodynamics}
\subsubsection{Independent field variables}
We begin by identifying the key results of the classical field theory which serve as the starting point of our canonical quantization scheme. The electrodynamics of continuous media in the absence of free charges is governed by the source-free macroscopic Maxwell equations, written here in Heaviside-Lorentz (H--L) units,
\begin{subequations}\begin{eqnarray}
\dot{\Dv}=c\;\curl\Hv,&\qquad&\dot{\Bv}=-c\;\curl\Ev,\label{curleqs}\\
\dive\Dv=0,&\qquad&\dive\Bv=0.\label{diveqs}
\end{eqnarray}\label{maxeqs}\end{subequations}
The field variables $\Dv(\rv,t)$, $\Bv(\rv,t)$, $\Ev(\rv,t)$ and $\Hv(\rv,t)$ are the electric induction, magnetic induction, electric field and magnetic field, respectively. A dot above a quantity denotes a time derivative, and $c$ is the speed of light in vacuo. Conversion to SI units is effected by replacement of $\Dv$, $\Bv$, $\Ev$, and $\Hv$, by $\Dv/\sqrt{\varepsilon_0}$, $\Bv/\sqrt{\mu_0}$, $\sqrt{\varepsilon_0}\;\Ev$, and $\sqrt{\mu_0}\;\Hv$, respectively, where $\varepsilon_0$ is the permittivity of free space and $\mu_0$ is the permeability of free space. Although the choice of H--L units is somewhat unorthodox, it is particularly convenient in the consideration of magneto-electric media in that the electromagnetic field variables all share the same dimensions. From the outset we may identify (\ref{diveqs}) as initial conditions for $\Dv$ and $\Bv$ since, on account of (\ref{curleqs}), if they are satisfied at one time they are satisfied at all times. In addition to (\ref{maxeqs}), a complete description of the field dynamics requires that one pair of the field variables be treated as independent and the remaining pair be expressed as functions of them through a set of constitutive relations. In doing so, there is a freedom regarding which two quantities are chosen as the independent pair. Here we take $\Dv$ and $\Bv$ to be the independent fields. In order to justify such a choice we write down the standard expression for the incremental change in the energy density as a result of changes in the fields alone,
\begin{equation}
{\rm d}\mathscr{U}=\Ev\cdot{\rm d}\Dv+\Hv\cdot{\rm d}\Bv.
\label{du}
\end{equation}
In the absence of dispersion the relationship between the field variables is local in time and $\Ev$ and $\Hv$ may then be written as
\begin{equation}
\Ev=\frac{\partial\mathscr{U}}{\partial\Dv},\qquad\Hv=\frac{\partial\mathscr{U}}{\partial\Bv},
\label{ehdef}
\end{equation}
which then allows (\ref{maxeqs}) to be rewritten solely in terms of $\Dv$, $\Bv$, and the energy density $\mathscr{U}$, viz.,
\begin{subequations}\begin{eqnarray}
\dot{\Dv}=c\;\curl\frac{\partial\mathscr{U}}{\partial\Bv},&\qquad&\dot{\Bv}=-c\;\curl\frac{\partial\mathscr{U}}{\partial\Dv},\label{curleqsdb}\\
\dive\Dv=0,&\qquad&\dive\Bv=0.\label{diveqsdb}
\end{eqnarray}\label{maxeqsdb}\end{subequations}
The pairs $(\Ev,\Bv)$ and $(\Dv,\Hv)$ are commonly viewed as primary and subsidiary variables, respectively (e.g., \cite{Sommerfeld:1952}). However, (\ref{du})--(\ref{maxeqsdb}) suggest that, for a Hamiltonian picture of dispersionless macroscopic electrodynamics in the absence of free charges, the natural choice of independent field variables is the pair $(\Dv,\Bv)$, as first noted by Born and Infeld \cite{Born:1934}. Motivated by this result, as well as the advantage of working with transverse fields, we extend this choice to the present treatment where dispersion and loss are included, and (\ref{ehdef}) no longer follows directly from (\ref{du}). Nonetheless, we note that the quantum analogue of (\ref{ehdef}) holds with the energy density $\mathscr{U}$ replaced with the Hamiltonian density corresponding to the Hamiltonian operator $\ham$ to be presented in Section \ref{hamiltoniansection}. 

The required constitutive relations must therefore express the fields $\Ev$ and $\Hv$ in terms of $\Dv$ and $\Bv$ at all times. With the standard definitions of the polarization $\Pv(\rv,t)$ and magnetization $\Mv(\rv,t)$ as
\begin{equation}
\Pv=\Dv-\Ev,\qquad\Mv=\Bv-\Hv,
\label{pmdef}
\end{equation}
this is equivalent to writing
\begin{equation}
\Pv=\Pv\left\{\Dv,\Bv\right\},\qquad\Mv=\Mv\left\{\Dv,\Bv\right\},
\label{pmdb}
\end{equation}
where the temporal non-locality is implied (SI units are obtained by the replacements $\Pv\to\Pv/\sqrt{\varepsilon_0}$ and $\Mv\to\sqrt{\mu_0}\;\Mv$). In the absence of explicit spatial dispersion, the dependencies of $\Pv$ and $\Mv$ upon the independent fields $\Dv$ and $\Bv$ in (\ref{pmdb}) represent the most general case of an anisotropic, gyrotropic, magneto-electric medium. The definitions of several other common classes of media are summarized in Table \ref{mediadefs}.
\begin{table}
\caption{\label{mediadefs}Definitions of some common classes of media through the dependencies of the polarization $\Pv$ and magnetization $\Mv$ upon the electric induction $\Dv$ and magnetic induction $\Bv$, when the latter pair are chosen as the independent fields.}
\begin{ruledtabular}
\begin{tabular}{ccc}
\multicolumn{1}{c}{Medium class} & $\Pv$ & $\Mv$ \\
\hline
\multicolumn{1}{l}{Dielectric} & $\Pv\{\Dv\}$ & $0$ \\
\multicolumn{1}{l}{Magnetic} & $0$ & $\Mv\{\Bv\}$ \\
\multicolumn{1}{l}{Magneto-dielectric} & $\Pv\{\Dv\}$ & $\Mv\{\Bv\}$ \\
\multicolumn{1}{l}{Magneto-electric} & $\Pv\{\Dv,\Bv\}$ & $\Mv\{\Dv,\Bv\}$ \\
\end{tabular}
\end{ruledtabular}
\end{table}

\subsubsection{\label{susceptibilitiessection}Susceptibility tensors}
The choice of independent fields, and therefore the form of the constitutive relations in (\ref{pmdb}), have ramifications for how the transfer of energy between the electromagnetic field and the medium is viewed. By considering this energy transfer we now derive some properties of the medium response functions. Standard manipulation of (\ref{curleqs}) yields
\begin{equation}
-\dive\Sv_{\rm EH}=\Ev\cdot\dot{\Dv}+\Hv\cdot\dot{\Bv},
\label{divs}
\end{equation}
where $\Sv_{\rm EH}=c\Ev\times\Hv$ is the Heaviside-Lorentz form of the electromagnetic energy flux associated with $\Ev$ and $\Hv$. With the present choice of independent variables it is natural to define the energy ${\rm U}_{\rm DB}$ associated with the fields $\Dv$ and $\Bv$ in the volume $\mathcal{V}$,
\begin{equation}
{\rm U}_{\rm DB}=\int_{\mathcal{V}}{\rm d}V\;\frac{1}{2}\left(\Dv\cdot\Dv+\Bv\cdot\Bv\right).
\label{udef}
\end{equation}
We then insert (\ref{pmdef}) into (\ref{divs}) and integrate over all time and over the volume $\mathcal{V}$ to obtain
\begin{equation}
-\intf\dt\oint_{\mathcal{S}}{\rm d}A\;\mathbf{n}\cdot\Sv_{\rm EH}=\Delta{\rm U}_{\rm DB}-\intf\dt\int_{\mathcal{V}}{\rm d}V\;\left(\Pv\cdot\dot{\Dv}+\Mv\cdot\dot{\Bv}\right),
\label{dsdef}
\end{equation}
where $\Delta{\rm U}_{\rm DB}$ is the total change in the independent field energy, and $\mathbf{n}$ is the outward unit vector normal to the surface $\mathcal{S}$. There has been much discussion regarding the correct definition of the Poynting vector as the representation of electromagnetic flux (e.g., \cite{McCall:2009}). By taking $\mathcal{S}$ to lie exclusively in the vacuum we avoid this controversy in that, for the purposes of the integral on the LHS of (\ref{dsdef}), all the possible flux vectors are equivalent; i.e., $\Sv_{\rm EH}=\Sv_{\rm EB}=\Sv_{\rm DH}=\Sv_{\rm DB}=c\Dv\times\Bv$. Assuming that the only flux of energy across $\mathcal{S}$ is of an electromagnetic nature, we may then use (\ref{dsdef}) to write the total change in the energy of the system enclosed within $\mathcal{V}$, $\Delta{\rm U}_{\rm tot}$, as
\begin{equation}
\Delta{\rm U}_{\rm tot}=-\intf\dt\oint_{\mathcal{S}}{\rm d}A\;\mathbf{n}\cdot\Sv_{\rm DB}=\Delta{\rm U}_{\rm DB}+\intf\dt\int_{\mathcal{V}}{\rm d}V\;\left(\dot{\Pv}\cdot\Dv+\dot{\Mv}\cdot\Bv\right).
\label{dudef}
\end{equation}
The second term on the RHS of (\ref{dudef}) represents the contribution to $\Delta{\rm U}_{\rm tot}$ due to the presence of a medium within $\mathcal{V}$. In order to relate this change in energy to the response of a magneto-electric medium, we write explicit expressions for the constitutive relations in (\ref{pmdb}) by defining the real susceptibility tensors $\Gv^{\si\nu}(\rv,t)$, with $\si,\nu={\rm e,m}$. These relate $\Pv$ and $\Mv$ at time $t$ to $\Dv$ and $\Bv$ at all times through
\begin{eqnarray}
\Pv(t)&=&\intf\dt\;\left[\Gvee(t-t')\cdot\Dv(t')+\Gvem(t-t')\cdot\Bv(t')\right],
\label{pvcrt}\\
\Mv(t)&=&\intf\dt\;\left[\Gvme(t-t')\cdot\Dv(t')+\Gvmm(t-t')\cdot\Bv(t')\right],
\label{mvcrt}
\end{eqnarray}
with the Fourier domain representation
\begin{eqnarray}
\Pv(\omega)&=&\Gvee(\omega)\cdot\Dv(\omega)+\Gvem(\omega)\cdot\Bv(\omega),
\label{pvcr}\\
\Mv(\omega)&=&\Gvme(\omega)\cdot\Dv(\omega)+\Gvmm(\omega)\cdot\Bv(\omega).
\label{mvcr}
\end{eqnarray}
In order to identify the dissipative and non-dissipative parts of the susceptibility tensors we define
\begin{eqnarray}
\Gvh^{\si\nu}(\omega)&=&\frac{1}{2}\left[\Gv^{\si\nu}(\omega)+\Gvb^{\nu\si*}(\omega)\right],
\label{ghdef}\\
\Gva^{\si\nu}(\omega)&=&\frac{1}{2}\left[\Gv^{\si\nu}(\omega)-\Gvb^{\nu\si*}(\omega)\right],
\label{gadef}
\end{eqnarray}
where an over-bar denotes a tensor with Cartesian components obtained from those of the unbarred quantity by a matrix transpose (recall that $\si,\nu={\rm e,m}$). Thus an over-bar combined with a star indicates the Hermitian transpose (the superscript $\dagger$ is reserved for the adjoint of an operator). This allows $\Gv^{\si\nu}$ to be written as
\begin{equation}
\Gv^{\si\nu}(\omega)=\Gvh^{\si\nu}(\omega)+\Gva^{\si\nu}(\omega).
\label{gvparts}
\end{equation}
We note that for $\si=\nu$, $\Gvh^{\si\si}(\omega)$ and $\Gva^{\si\si}(\omega)$ are the Hermitian and anti-Hermitian parts, respectively, of $\Gv^{\si\si}(\omega)$. We may then employ (\ref{pvcr})--(\ref{gadef}) to express the second term on the RHS in (\ref{dudef}) as
\begin{eqnarray}
&&\intf\dt\int_{\mathcal{V}}{\rm d}V\;\left(\dot{\Pv}\cdot\Dv+\dot{\Mv}\cdot\Bv\right)\nonumber\\
&=&2\inth\frac{{\rm d}\omega}{2\pi}\;\left(-i\omega\right)\intV\left\{
\left[\begin{array}{cc}
\Dv^*(\omega) & \Bv^*(\omega)
\end{array}\right]
\left[\begin{array}{cc}
\Gva^{\rm ee}(\omega) & \Gva^{\rm em}(\omega) \\
\Gva^{\rm me}(\omega) & \Gva^{\rm mm}(\omega)
\end{array}\right]
\left[\begin{array}{c}
\Dv(\omega) \\ \Bv(\omega)
\end{array}\right]\right\},
\label{diss}
\end{eqnarray}
where the Fourier transform of an arbitrary function $f(t)$ is defined as
\begin{equation}
f(\omega)=\int^\infty_{-\infty}{\rm d}t\;e^{i\omega t}f(t).
\label{ftdef}
\end{equation}
Now consider a transient interaction of the electromagnetic field with a medium of finite extent, such that $\mathcal{S}$ may be assumed to lie in the vacuum and $\Delta{\rm U}_{\rm DB}=0$ (${\rm U}_{\rm DB}(t)=0$ for $t=\pm\infty$, say). From (\ref{dudef}) we see that the RHS of (\ref{diss}) is unambiguously the total energy transferred between the electromagnetic field and the medium. Thus, only the medium response described by $\Gva^{\si\nu}(\omega)$ leads to gain or dissipation of electromagnetic energy, and $\Gvh^{\si\nu}(\omega)$ describes the non-dissipative interaction.

Further symmetries of the susceptibility tensors may be determined through applying time reversal. We impose the following properties upon the fields in the time domain:
\begin{eqnarray}
&&\left\{\Dv\right\}_{-\mathscr{T}}=\Dv,\qquad\left\{\Bv\right\}_{-\mathscr{T}}=-\Bv,\nonumber\\
&&\left\{\Pv\right\}_{-\mathscr{T}}=\Pv,\qquad\left\{\Mv\right\}_{-\mathscr{T}}=-\Mv,
\label{trdbpm}
\end{eqnarray}
where $\left\{\right\}_{-\mathscr{T}}$ represents the operation of time-reversal upon the enclosed expression. Additionally, the application of time-reversal in the Fourier domain corresponds to the replacement $\omega\to-\omega$, as well as the appropriate transformation of any parameters; e.g., in the case where the susceptibility tensors are dependent upon an ambient magnetic field $\Bv_0$ we have $\left\{\Gva^{\si\nu}(\omega,\Bv_0)\right\}_{-\mathscr{T}}=\left\{\Gva^{\si\nu}(-\omega,\Bv_0)\right\}_{-\mathbf{B}_0}=\Gva^{\si\nu}(-\omega,-\Bv_0)$. Using $\{\}_{-\mathbf{B}_0}$ to represent the time reversal of all such parameters, we may then use (\ref{diss}), along with the reality condition for the fields, to obtain
\begin{equation}
\left\{\Gva^{\si\si}(\omega)\right\}_{-\mathbf{B}_0}=\Gvba^{\si\si}(\omega),\qquad\left\{\Gva^{\si\nu}(\omega)\right\}_{-\mathbf{B}_0}=-\Gvba^{\nu\si}(\omega),\qquad\si\ne\nu.
\label{gaor}
\end{equation}
Expressions analogous to (\ref{gaor}) for $\Gvh^{\si\nu}(\omega)$ are obtained by imposing causality. This amounts to setting $\Gv^{\si\nu}(t)=0$ for $t<0$, which results in the Kramers-Kronig relations \cite{Landau:1980a,Melrose:1991}
\begin{eqnarray}
\Gvh^{\si\nu}(\omega)&=&2i\mathscr{P}\intf\frac{{\rm d}\omega'}{2\pi}\;\frac{\Gva^{\si\nu}(\omega')}{\omega-\omega'},
\label{ghkk}\\
\Gva^{\si\nu}(\omega)&=&2i\mathscr{P}\intf\frac{{\rm d}\omega'}{2\pi}\;\frac{\Gvh^{\si\nu}(\omega')}{\omega-\omega'},
\label{gakk}
\end{eqnarray}
where $\mathscr{P}$ indicates a Cauchy principal value integral. Combining (\ref{gaor})--(\ref{gakk}) then leads to
\begin{equation}
\left\{\Gvh^{\si\si}(\omega)\right\}_{-\mathbf{B}_0}=\Gvbh^{\si\si}(\omega),\qquad\left\{\Gvh^{\si\nu}(\omega)\right\}_{-\mathbf{B}_0}=-\Gvbh^{\nu\si}(\omega),\qquad\si\ne\nu.
\label{ghor}
\end{equation}
From (\ref{gvparts}), (\ref{gaor}), and (\ref{ghor}) we may now identify the Onsager relations
\begin{equation}
\left\{\Gv^{\si\si}\right\}_{-\mathbf{B}_0}=\Gvb^{\si\si},\qquad\left\{\Gv^{\si\nu}\right\}_{-\mathbf{B}_0}=-\Gvb^{\nu\si},\qquad\si\ne\nu,
\label{gor}
\end{equation}
which hold in both the time and frequency domains. It should be noted that in deriving the Kramers-Kronig and Onsager relations expressed in (\ref{ghkk}), (\ref{gakk}), and (\ref{gor}), only the macroscopic Maxwell equations, the behaviour of the fields under time reversal, and the assumption of causality have been used. Therefore, these conditions upon the susceptibility tensors represent fundamental properties of a causal, electromagnetic  medium, and they must be reflected in a valid quantum theory.

\subsection{\label{hamiltoniansection}Hamiltonian operator}
We now turn to the construction of a Hamiltonian operator that, with the standard commutation relations, leads to the quantum analogue of (\ref{maxeqs}) with constitutive relations of the form given in (\ref{pvcrt}) and (\ref{mvcrt}). The associated susceptibility tensors must satisfy standard Kramers-Kronig and Onsager relations. Some Hamiltonian formulations of macroscopic electrodynamics have proceeded from a Lagrangian \cite{Huttner:1992,Suttorp:2004,Philbin:2010}. However, it is sufficient to provide a Hamiltonian operator directly. Indeed, transforming from the Lagrangian to a Hamiltonian leads to a complicated field-medium interaction that is difficult to diagonalize. Instead we directly construct a Hamiltonian which describes a rather general class of magneto-electric media. This allows us to define the various couplings straightforwardly in terms of the canonical variables, and avoid the complications of transitioning from a Lagrangian to a Hamiltonian picture. In addition, the classical value of the Hamiltonian must be equal to the energy of the system, which is satisfied in our theory below by construction. The full system consists of two linearly coupled subsystems representing the vacuum electromagnetic field and a medium. The corresponding Hamiltonian is therefore of the form
\begin{equation}
\ham=\ham_{\rm emf}+\ham_{\rm med}+\ham_{\rm int},
\label{htot}
\end{equation}
where $\ham_{\rm emf}$, $\ham_{\rm med}$, and $\ham_{\rm int}$ are the electromagnetic field, medium, and interaction Hamiltonians, respectively, and a hat denotes an operator. In what follows, all operators may be presumed to commute unless specified otherwise. In the Heisenberg picture, the time evolution of an arbitrary operator $\hat{O}$ is governed by the equation
\begin{equation}
i\hbar\;\dot{\hat{O}}=\left[\hat{O},\ham\right],
\label{hbeq}
\end{equation}
where $[\;,\;]$ represents a commutator, and the dot denotes a total derivative with respect to time (in that the components of the field point $\rv$ are not dynamical variables).

The Hamiltonian for the electromagnetic field is \cite{Born:1934} (c.f. (\ref{udef}))
\begin{equation}
\ham_{\rm emf}=\frac{1}{2}\int\dV\;\left[\Dop(\rv,t)\cdot\Dop(\rv,t)+\Bop(\rv,t)\cdot\Bop(\rv,t)\right],
\label{hvac}
\end{equation}
and the equal time commutation relations (ETCRs) for the components of the field operators are \cite{Heisenberg:1929,Born:1934}
\begin{equation}
\left[\hat{D}_i(\rv,t),\hat{B}_j(\rv',t)\right]=i\hbar c\epsilon_{ikj}\frac{\partial}{\partial r_k}\delta(\rv-\rv'),
\label{dbcr}
\end{equation}
where Cartesian vector and tensor components are indexed by Latin subscripts, a sum over repeated indices is implied, $\boldsymbol{\epsilon}$ is the Levi-Civita pseudotensor, and $\delta(\rv)$ is the Dirac delta function.

The construction of the medium and interaction Hamiltonians requires some discussion. Previous models for a medium which exhibits both an electric and magnetic response have typically involved independent electric and magnetic subsystems. In the application of the phenomenological approach to a magneto-dielectric medium \cite{Dung:2003}, the noise polarization and magnetization are made to originate from independent bosonic vector and pseudovector fields, respectively. Such a separation of the medium is deemed appropriate in modeling materials where the electric and magnetic responses arise from physically distinct material constituents or degrees of freedom. Similarly, in the canonical treatment of a magneto-dielectric \cite{Philbin:2010} this prescription is reflected in the introduction of two separate sets of harmonic oscillator fields in the model for the medium. Following on from this work, the identification of these fields as corresponding to electric and magnetic oscillators was then made explicitly in the construction of a Lagrangian for a magneto-electric medium \cite{Horsley:2011}. This was effected by associating the symmetry properties of the oscillator amplitudes under spatial inversion and time reversal with those of the electric or magnetic field, as appropriate. We have found that a magneto-electric response may be obtained with a single set of oscillators. However, such a response is severely restricted (e.g., the cross-coupling is completely fixed by the dielectric and magnetic responses in the isotropic case). Here we present the most general form of magneto-electric response obtainable within the context of the established harmonic oscillator model. Since the coupling of the field to multiple continua is already implied in the vector nature of the oscillators, it is a straightforward matter to add additional degrees of freedom to the medium. We pursue this by including an arbitrary number of such oscillator sets (to represent a MM with electromagnetic resonances associated with several material constituents, for instance). In general, we may expect each class of oscillators to exhibit a magneto-electric coupling, with no purely electric or magnetic character, and we model the medium with $N$ sets of \emph{vector} operators $\bop_{\la\om}(\rv,t)$, $\la=1,\ldots ,N$, representing bosonic excitations associated with each frequency $\om>0$ and field point $\rv$. The required symmetry properties of the response are then imposed on the coupling coefficients between the field and medium. The corresponding Hamiltonian for the medium is
\begin{equation}
\ham_{\rm med}=\sum_\la\intV\inth\dom\;\hbar\om\;\bopa_{\la\om}(\rv,t)\cdot\bop_{\la\om}(\rv,t),
\label{hmed}
\end{equation}
and the vector components of the medium operators obey the ETCR
\begin{equation}
\left[\hat{b}_{\la\om i}(\rv,t),\hat{b}^{\dagger}_{\la'\om'j}(\rv',t)\right]=2\pi\delta_{\la\la'}\delta_{ij}\;\delta(\om-\om')\;\delta(\rv-\rv').
\label{bbcr}
\end{equation}
where $\delta_{\la\la'}$ and $\delta_{ij}$ are Kronecker deltas. The interaction Hamiltonian is then constructed as a spatially and temporally local, bilinear coupling between the vacuum electromagnetic field and medium operators, viz.,
\begin{eqnarray}
\ham_{\rm int}&=&-\hbsr\intV\;\Dop(\rv,t)\cdot\sum_\la\inth\dom\;\left[\Lve_\la(\rv,\om)\cdot\bop_{\la\om}(\rv,t)+\Lv^{\rm e*}_\la(\rv,\om)\cdot\bopa_{\la\om}(\rv,t)\right]\nonumber\\
&&-\hbsr\intV\;\Bop(\rv,t)\cdot\sum_\la\inth\dom\;\left[\Lvm_\la(\rv,\om)\cdot\bop_{\la\om}(\rv,t)+\Lv^{\rm m*}_\la(\rv,\om)\cdot\bopa_{\la\om}(\rv,t)\right],\nonumber\\
\label{hint}
\end{eqnarray}
where the complex valued, second rank proper tensors $\Lve_\la(\rv,\om)$ and pseudotensors $\Lvm_\la(\rv,\om)$ are defined for positive $\om$. Finally, we assume that the microscopic dynamics that underlie our macroscopic picture lead to a Hamiltonian which is symmetric under time reversal. We therefore require that $\Lve_\la\to\Lv^{\rm e*}_\la$ and $\Lvm_\la\to-\Lv^{\rm m*}_\la$ under this operation. This requirement also ensures consistency with the classical theory (see (\ref{podef}) and (\ref{modef}) below). For the alternate representation where the medium fields are represented as harmonic oscillators with coordinate operators $\hat{\boldsymbol{q}}_{\la\om}(\rv,t)$ and conjugate momenta $\hat{\boldsymbol{p}}_{\la\om}(\rv,t)$ (e.g., \cite{Bhat:2006,Philbin:2010}) the form of (\ref{hint}) corresponds to the coupling of $\Dop$ and $\Bop$ to both $\hat{\boldsymbol{q}}_{\la\om}$ and $\hat{\boldsymbol{p}}_{\la\om}$. Ultimately, the coupling tensors $\Lve_\la$ and $\Lvm_\la$ in (\ref{hint}) are to be determined by the measured or calculated susceptibility of the medium. However, making the replacements $\Lv^\si_\la\to\Lv^\si_\la\cdot\Uvb_\la^*$ and $\bop_{\la\om}\to\Uv_\la\cdot\bop_{\la\om}$, where $\si={\rm e,m}$ and the tensor $\Uv_\la$ represents an arbitrary unitary transformation, leaves the Hamiltonian unchanged. This represents an inherent freedom in the model for the medium.

Thus, inserting (\ref{hvac}), (\ref{hmed}), and (\ref{hint}) into (\ref{htot}) we may write the full Hamiltonian explicitly as
\begin{eqnarray}
\ham&=&\frac{1}{2}\int\dV\;\left[\Dop(\rv,t)\cdot\Dop(\rv,t)+\Bop(\rv,t)\cdot\Bop(\rv,t)\right]\nonumber\\
&&+\sum_\la\intV\inth\dom\;\hbar\om\;\bopa_{\la\om}(\rv,t)\cdot\bop_{\la\om}(\rv,t)\nonumber\\
&&-\hbsr\intV\;\Dop(\rv,t)\cdot\sum_\la\inth\dom\;\left[\Lve_\la(\rv,\om)\cdot\bop_{\la\om}(\rv,t)+\Lv^{\rm e*}_\la(\rv,\om)\cdot\bopa_{\la\om}(\rv,t)\right]\nonumber\\
&&-\hbsr\intV\;\Bop(\rv,t)\cdot\sum_\la\inth\dom\;\left[\Lvm_\la(\rv,\om)\cdot\bop_{\la\om}(\rv,t)+\Lv^{\rm m*}_\la(\rv,\om)\cdot\bopa_{\la\om}(\rv,t)\right].
\label{htotexp}
\end{eqnarray}

\subsection{\label{qmesection}Dynamical equations and constitutive relations}
To demonstrate the consistency of the Hamiltonian system presented in Section \ref{hamiltoniansection} with macroscopic electromagnetism, we first insert $\Dop$ and $\Bop$ into (\ref{hbeq}) and use (\ref{dbcr}) to obtain
\begin{equation}
\dot{\Dop}=c\curl\Bop-c\curl\hbsr\sum_\la\inth\dom\;\left[\Lvm_\la(\om)\cdot\bop_{\la\om}+\Lv^{\rm m*}_\la(\om)\cdot\bopa_{\la\om}\right],
\label{ddot}
\end{equation}
and
\begin{equation}
\dot{\Bop}=-c\curl\Dop+c\curl\hbsr\sum_\la\inth\dom\;\left[\Lve_\la(\om)\cdot\bop_{\la\om}+\Lv^{\rm e*}_\la(\om)\cdot\bopa_{\la\om}\right].
\label{bdot}
\end{equation}
Recalling (\ref{ehdef}) and making the operator definitions analogous to (\ref{pmdef}),
\begin{equation}
\Pop=\Dop-\Eop,\qquad\Mop=\Bop-\Hop,
\label{pomodef}
\end{equation}
we are led to identify the polarization $\Pop(\rv,t)$ and magnetization $\Mop(\rv,t)$ of the medium as
\begin{eqnarray}
\Pop(\rv,t)&=&\hbsr\sum_\la\inth\dom\;\left[\Lve_\la(\rv,\om)\cdot\bop_{\la\om}(\rv,t)+\Lv^{\rm e*}_\la(\rv,\om)\cdot\bopa_{\la\om}(\rv,t)\right],
\label{podef}\\
\Mop(\rv,t)&=&\hbsr\sum_\la\inth\dom\;\left[\Lvm_\la(\rv,\om)\cdot\bop_{\la\om}(\rv,t)+\Lv^{\rm m*}_\la(\rv,\om)\cdot\bopa_{\la\om}(\rv,t)\right].
\label{modef}
\end{eqnarray}
Note that consistency of (\ref{podef}) and (\ref{modef}) with (\ref{trdbpm}) follows from the time reversal properties imposed upon the coupling tensors. Inserting (\ref{pomodef})--(\ref{modef}) into (\ref{ddot}) and (\ref{bdot}) we obtain the quantum analogue of the Maxwell curl equations, (\ref{curleqs}), as
\begin{equation}
\dot{\Dop}=c\;\curl\Hop,\qquad\dot{\Bop}=-c\;\curl\Eop.
\label{qcurleqs}
\end{equation}
Additionally, the conditions
\begin{equation}
\dive\Dop=0,\qquad\dive\Bop=0,
\label{qdiveqs}
\end{equation}
must be enforced independently at this stage to reproduce the full set of macroscopic Maxwell equations. The description of the dynamics of the electromagnetic field operators contained within (\ref{pomodef})--(\ref{qdiveqs}) remains incomplete, however, until we establish constitutive relations of the form (\ref{pvcrt}) and (\ref{mvcrt}).

To identify the required constitutive relations for the medium we must determine the contributions to the dynamics of $\bop_{\la\om}(\rv,t)$ driven by the electromagnetic field. Having thus expressed the medium operators in terms of $\Dop$ and $\Bop$, we may substitute the results into (\ref{podef}) and (\ref{modef}) to yield the field-induced polarization and magnetization. The dynamical equation for the medium operators follows from (\ref{hbeq}), (\ref{bbcr}), and (\ref{htotexp}), as
\begin{equation}
\dot{\bop}_{\la\om}(\rv,t)=-i\om\bop_{\la\om}(\rv,t)+i\hbsri\left[\Lvb^{\rm e*}_\la(\rv,\om)\cdot\Dop(\rv,t)+\Lvb^{\rm m*}_\la(\rv,\om)\cdot\Bop(\rv,t)\right].
\label{bmdot}
\end{equation}
Integrating this equation directly for some initial time $\tau<t$ we obtain
\begin{eqnarray}
\bop_{\la\om}(\rv,t)&=&i\hbsri\int^\infty_\tau\dt'\;\theta(t-t')\;e^{-i\om(t-t')}\left[\Lvb^{\rm e*}_\la(\rv,\om)\cdot\Dop(\rv,t')+\Lvb^{\rm m*}_\la(\rv,\om)\cdot\Bop(\rv,t')\right]\nonumber\\
&&+\bop_{\la\om}(\rv,\tau)\;e^{-i\om(t-\tau)}.
\label{bsol}
\end{eqnarray}
where the causality in the relationship between the electromagnetic field operators $\Dop$ and $\Bop$, and the medium operators $\bop_{\la\om}$ is clear. The use of the Heaviside step function $\theta(t)$ has allowed extension of the upper limit of the integral to $+\infty$.

Returning to the task of identifying the constitutive relations, we substitute (\ref{bsol}) in (\ref{podef}) and (\ref{modef}) to obtain expressions for the polarization and magnetization operators,
\begin{eqnarray}
\Pop(\rv,t)&=&\int^{\infty}_{-\infty}\dt'\;\left[\Gvee(\rv,t-t')\cdot\Dop(\rv,t')+\Gvem(\rv,t-t')\cdot\Bop(\rv,t')\right]\nonumber\\
&&+\Pop^{\rm (n)}(\rv,t),
\label{podefg}\\
\Mop(\rv,t)&=&\int^{\infty}_{-\infty}\dt'\;\left[\Gvme(\rv,t-t')\cdot\Dop(\rv,t')+\Gvmm(\rv,t-t')\cdot\Bop(\rv,t')\right]\nonumber\\
&&+\Mop^{\rm (n)}(\rv,t),
\label{modefg}
\end{eqnarray}
where the $\Gv^{\si\nu}$'s are identified below, and we have defined the noise operators
\begin{eqnarray}
\Pop^{\rm (n)}(\rv,t)&=&\hbsr\sum_\la\inth\dom\;\left[\Lve_\la(\rv,\om)\cdot\bop_{\la\om}(\rv,\tau)\;e^{-i\om(t-\tau)}\right..\nonumber\\
&&\hspace{5cm}\left.+\Lv^{\rm e*}_\la(\rv,\om)\cdot\bopa_{\la\om}(\rv,\tau)\;e^{i\om(t-\tau)}\right],
\label{pondeft}\\
\Mop^{\rm (n)}(\rv,t)&=&\hbsr\sum_\la\inth\dom\;\left[\Lvm_\la(\rv,\om)\cdot\bop_{\la\om}(\rv,\tau)\;e^{-i\om(t-\tau)}\right.\nonumber\\
&&\hspace{5cm}\left.+\Lv^{\rm m*}_\la(\rv,\om)\cdot\bopa_{\la\om}(\rv,\tau)\;e^{i\om(t-\tau)}\right].
\label{mondeft}
\end{eqnarray}
The domains of the integrals in (\ref{podefg}) and (\ref{modefg}) have been extended to $-\infty$ by choosing $\tau$ such that $t-\tau>\tau_R$, where $\tau_R$ is the finite response time of the medium. The expressions (\ref{podefg}) and (\ref{modefg}) form the quantum analogue of the constitutive relations (\ref{pvcrt}) and (\ref{mvcrt}), as desired, with the addition of the noise terms $\Pop^{\rm (n)}(\rv,t)$ and $\Mop^{\rm (n)}(\rv,t)$ involving the initial conditions for the $\bop_{\la\om}(\rv,t)$ operators. Such noise operators are a hallmark of dissipative quantum systems wherein they act to preserve the commutation relations by compensating for the otherwise dissipative decay of the coupled operators.

The causal susceptibility tensors $\Gv^{\si\nu}(\rv,t)$ in (\ref{podefg}) and (\ref{modefg}) are related to the coupling tensors $\Lv^\si_\la(\om,\rv)$ ($\si,\nu={\rm e,m}$) through
\begin{eqnarray}
\Gv^{\si\nu}(\rv,t)&=&\theta(t)\sum_\la\inth\dom\;\left[2\mathscr{R}\left\{\Lv_\la^{\si}(\rv,\om)\cdot\Lvb_\la^{\nu*}(\rv,\om)\right\}\;\sin(\om t)\right.\nonumber\\
&&\hspace{3.5cm}-\left.2\mathscr{I}\left\{\Lv_\la^{\si}(\rv,\om)\cdot\Lvb_\la^{\nu*}(\rv,\om)\right\}\;\cos(\om t)\right],
\label{gdef}
\end{eqnarray}
where the symbols $\mathscr{R}\{\}$ and $\mathscr{I}\{\}$ represent the real and imaginary parts of the enclosed expressions, respectively. The Heaviside step function on the RHS of (\ref{gdef}), along with the time reversal properties imposed upon the $\Lv^\si_\la$'s, ensures that the identification made in equating the LHS and RHS of (\ref{gdef}) is consistent with the requirement that the $\Gv^{\si\nu}$'s satisfy the Kramers-Kronig and Onsager relations as expressed in (\ref{ghkk}), (\ref{gakk}), and (\ref{gor}). In the Fourier domain, $\Gv^{\si\nu}(\rv,\omega)$ may be separated as in (\ref{gvparts}), and by taking the Fourier transform of (\ref{gdef}) we may identify $\Gvh^{\si\nu}(\rv,\omega)$ and $\Gva^{\si\nu}(\rv,\omega)$, defined as in (\ref{ghdef}) and (\ref{gadef}), respectively, in terms of the coupling tensors as
\begin{equation}
\Gvh^{\si\nu}(\rv,\omega)=\sum_\la\mathscr{P}\inth\dom\;\frac{2\om\;\mathscr{R}\left\{\Lv_\la^\si(\rv,\om)\cdot\Lvb_\la^{\nu*}(\rv,\om)\right\}+2i\omega\mathscr{I}\left\{\Lv^\si_\la(\rv,\om)\cdot\Lvb_\la^{\nu*}(\rv,\om)\right\}}{\om^2-\omega^2},
\label{ghldef}
\end{equation}
and
\begin{equation}
\Gva^{\si\nu}(\rv,\omega)=\frac{i}{2}\frac{\omega}{|\omega|}\sum_\la\Lv^\si_\la(\rv,|\omega|)\cdot\Lvb_\la^{\nu*}(\rv,|\omega|).
\label{galdef}
\end{equation}
Thus, (\ref{podefg})--(\ref{galdef}) demonstrate how the temporally local coupling in (\ref{hint}) is related to the causal response of the medium.

So far we have demonstrated that the Hamiltonian and accompanying ETCR's introduced in Section \ref{hamiltoniansection} lead to the quantum analogue of the macroscopic Maxwell equations with magneto-electric constitutive relations involving susceptibility tensors which obey the Kramers-Kronig and Onsager relations. In addition, expressions for the non-classical noise operators $\Pop^{\rm (n)}$ and $\Mop^{\rm (n)}$ have been obtained, which fully determine their commutation relations and time evolution properties through dependence upon the initial conditions for the medium operators in the form $\bop_{\la\om}(\rv,\tau)\;e^{-i\om(t-\tau)}$. Thus, Section \ref{quantsection} constitutes a canonical quantization of macroscopic electrodynamics in a causal magneto-electric medium, within the restrictions placed upon the susceptibility tensors by their relationship to the coupling tensors in (\ref{gdef})--(\ref{galdef}) (see discussion in Section \ref{discussionsection}).

\section{\label{polaritonsection}Polariton operators}
The quantization of Section \ref{quantsection} has provided us with the dynamical equations for the electromagnetic field operators, and we are thus in a position to identify their solutions. The resulting expressions for $\Dop$, $\Bop$, $\Eop$, and $\Hop$ constitute the main results of the present work and are derived below in the form of the mode expansions (\ref{doex}), (\ref{boex}), (\ref{eoex}) and (\ref{hoex}), with the mode fields following from solutions to (\ref{ghep}). In obtaining these solutions we employ standard methods of modern quantum optics, which we now proceed to describe.

In the Heisenberg picture the dynamics of coupled quantum systems are often simplified by determining the eigen-operators of the full Hamiltonian which exhibit harmonic time dependence. When associated with dissipative systems, the application of this procedure is known as Fano theory \cite{Fano:1961,Barnett:1997}. In the present context of macroscopic electrodynamics we term the eigen-operators of $\ham$ as the \emph{polariton operators} which are required to satisfy the eigen-equation
\begin{equation}
\hbar\om\;\hat{O}_\om(t)=\left[\hat{O}_\om(t),\ham\right],
\label{oee}
\end{equation}
whence, with (\ref{hbeq}), $\hat{O}_\om(t)=\hat{O}_\om(0)\exp(-i\om t)$ follows. However, (\ref{oee}) tells us nothing about how the spatial degrees of freedom in the Hilbert space implied by (\ref{htotexp}) are to be accounted for in the labeling of the polariton operators.

In the canonical quantization of the vacuum electromagnetic field the concept of photons arises in connection with the plane wave modes of the vacuum. These photon modes are associated with the purely time dependent eigen-operators of the vacuum field Hamiltonian $\ham_{\rm emf}$, which evolve as $\aop_{u\mathbf{k}}(t)=\aop_{u\mathbf{k}}\exp(-i\omega_{\mathbf{k}}t)$. This allows the electromagnetic field operators to be written as expansions of the form \cite{Landau:1982}
\begin{eqnarray}
\Dop(\rv,t)&=&\sum_{u,\mathbf{k}}\left[\aop_{u\mathbf{k}}(t)\;\Dv_{u\mathbf{k}}(\rv)+\aopa_{u\mathbf{k}}(t)\;\Dv^*_{u\mathbf{k}}(\rv)\right],
\label{dvex}\\
\Bop(\rv,t)&=&\sum_{u,\mathbf{k}}\left[\aop_{u\mathbf{k}}(t)\;\Bv_{u\mathbf{k}}(\rv)+\aopa_{u\mathbf{k}}(t)\;\Bv^*_{u\mathbf{k}}(\rv)\right],
\label{bvex}
\end{eqnarray}
where $\Dv_{u\mathbf{k}}(\rv)$ and $\Bv_{u\mathbf{k}}(\rv)$ are the vacuum wave modes with wavevector $\mathbf{k}$ and polarization index $u=1,2$. The discrete sum over $\mathbf{k}$ follows from normalization to a finite box with periodic boundary conditions. In the extension of this approach to a linear, inhomogeneous, but non-dispersive medium (e.g., \cite{Bhat:2006}), the forms of (\ref{dvex}) and (\ref{bvex}) are preserved with the plane wave solutions being replaced with the electromagnetic modes of the structured medium. Such a description is particularly appropriate in the context of guided-wave optics and photonics, where the spatial modes of waveguiding structures and optical cavities form the natural language for the dynamics of the system. For this purpose it is desirable to maintain the modal approach when extending the quantum treatment of the electromagnetic field to Kramers-Kronig media, as is done here. We therefore introduce modal polariton operators in order to build a description of the electromagnetic field operators $\Dop$ and $\Bop$ analogous to (\ref{dvex}) and (\ref{bvex}). The polariton operators are labeled by the indices $n$ and $\om$, the nature of which follows from consideration of the Hilbert space implied by the form of (\ref{htotexp}): the index $\om$ is continuous, while the precise nature of the index $n$, which represents the spatial degrees of freedom, is determined by the geometry of the system along with the boundary conditions imposed: ultimately we shall find that $n$ labels spatial electromagnetic field distributions corresponding to the polariton modes in analogy with the labels $(u,\mathbf{k})$ for the photon modes. For convenience, as in the vacuum mode case, we assume normalization within a finite box with periodic boundary conditions and therefore treat $n$ as a discrete index; the generalization of this prescription to multiple or continuous indices is straightforward and may be made later as required.

\subsection{\label{tplpsection}Transverse and longitudinal response polaritons}
From the outset we partition the modal polaritons into two classes: transverse response polaritons (TP's) and longitudinal response polaritons (LP's), leading to the Hamiltonian form
\begin{equation}
\ham=\sum_n\inth\dom\;\hbar\om\;\copa_{\om n}(t)\;\cop_{\om n}(t)+
\sum_\la\sum_{n}\inth\dom\;\hbar\om\;\sopa_{\la\om n}(t)\;\sop_{\la\om n}(t),
\label{nlpham}
\end{equation}
where $\cop_{\om n}(t)$ and $\sop_{\la\om n}(t)$ correspond to the TP and LP operators, respectively. The TP operators represent  the collective field-medium excitations and form a single class. The LP's represent excitations of the medium which do not couple to the transverse electromagnetic field and therefore constitute at most $N$ subclasses for each subsystem of the medium. By assumption, the polariton operators satisfy the equations
\begin{eqnarray}
\hbar\om\;\cop_{\om n}(t)&=&\left[\cop_{\om n}(t),\ham\right],\label{cee}\\
\hbar\om\;\sop_{\la\om n}(t)&=&\left[\sop_{\la\om n}(t),\ham\right],\label{see}
\end{eqnarray}
which imply $\cop_{\om n}(t)=\cop_{\om n}\;e^{-i\om t}$ and $\sop_{\la\om n}(t)=\sop_{\la\om n}\;e^{-i\om t}$. We also impose the ETCR's
\begin{eqnarray}
\left[\cop_{\om n},\copa_{\om'n'}\right]&=&2\pi\delta_{nn'}\delta(\om-\om'),\label{cccr}\\
\left[\sop_{\la\om n},\sopa_{\la'\om'n'}\right]&=&2\pi\delta_{\la\la'}\delta_{nn'}\delta(\om-\om'),\label{sscr}\\
\left[\cop_{\om n},\sopa_{\la\om'n'}\right]&=&0.\label{cscr}
\end{eqnarray}
The last of these ETCR's establishes the formal separation of the two classes of polariton operators, the classification of which follows from the definition of the LP modes as those associated with configurations of the medium which do not interact with the transverse $\Dop$ and $\Bop$ fields through the interaction Hamiltonian. This may be expressed in the form of the condition
\begin{equation}
\left[\sop_{\la\om n},\ham_{\rm int}^{(\la)}\right]=0,
\label{sdef}
\end{equation}
where
\begin{eqnarray}
\ham_{\rm int}^{(\la)}&=&-\hbsr\intV\;\Dop(\rv,t)\cdot\inth\dom\;\left[\Lve_\la(\rv,\om)\cdot\bop_{\la\om}(\rv,t)+\Lv^{\rm e*}_\la(\rv,\om)\cdot\bopa_{\la\om}(\rv,t)\right]\nonumber\\
&&-\hbsr\intV\;\Bop(\rv,t)\cdot\inth\dom\;\left[\Lvm_\la(\rv,\om)\cdot\bop_{\la\om}(\rv,t)+\Lv^{\rm m*}_\la(\rv,\om)\cdot\bopa_{\la\om}(\rv,t)\right].
\label{hintldef}
\end{eqnarray}
The partitioning of the polaritons into TP's and LP's combined with the transversality of the electromagnetic field operators now allows us to derive a number of useful results.

Since the polariton operators are the eigen-operators of the full Hamiltonian we may write $\Dop$, $\Bop$,  and $\bop$ as expansions of the form
\begin{eqnarray}
\Dop(\rv,t)&=&\sum_n\inth\dom\;\left[\cop_{\om n}(t)\;\Dv_{\om n}(\rv)+\copa_{\om n}(t)\;\Dv^*_{\om n}(\rv)\right],\label{doex}\\
\Bop(\rv,t)&=&\sum_n\inth\dom\;\left[\cop_{\om n}(t)\;\Bv_{\om n}(\rv)+\copa_{\om n}(t)\;\Bv^*_{\om n}(\rv)\right],\label{boex}\\
\bop_{\la\om}(\rv,t)&=&\sum_n\inth\dom'\;\left[\cop_{\om'n}(t)\;\av_{\la n}(\rv,\om',\om)+\copa_{\om'n}(t)\;\bv^*_{\la n}(\rv,\om',\om)\right]\nonumber\\
&&+\sum_n\sop_{\la\om n}(t)\;\rhv_{\la n}(\rv,\om),\label{bex}
\end{eqnarray}
where $\Dv_{\om n}(\rv)$, $\Bv_{\om n}(\rv)$, $\av_{\la n}(\rv,\om',\om)$, $\bv_{\la n}(\rv,\om',\om)$, and $\rhv_{\la n}(\rv,\om)$ are vector field coefficients.

To obtain the inverse transformations corresponding to (\ref{doex})--(\ref{bex}) we construct nominal expansions of the polariton operators in terms of the subsystem operators $\Dop$, $\Bop$, and $\bop$. Using these expansions to evaluate the ETCR's between the subsystem and polariton operators (e.g., $[\Dop(\rv),\cop_{\om n}]$) and comparing the results with the same ETCR's evaluated using (\ref{doex})--(\ref{bex}), we obtain
\begin{eqnarray}
\cop_{\om n}(t)&=&\frac{1}{\hbar\om}\int\dV\;\left(\Evtc_{\om n}(\rv)\cdot\Dop(\rv,t)+\Hvtc_{\om n}(\rv)\cdot\Bop(\rv,t)\right)\nonumber\\
&&+\sum_\la\int\dV\inth\dom'\left[\av^*_{\la n}(\rv,\om,\om')\cdot\bop_{\la\om'}(\rv,t)-\bv^*_{\la n}(\rv,\om,\om')\cdot\bopa_{\la\om'}(\rv,t)\right],\nonumber\\
\label{cex}\\
\sop_{\la\om n}(t)&=&\int\dV\;\rhv^*_{\la n}(\rv,\om)\cdot\bop_{\la\om}(\rv,t),
\label{sex}
\end{eqnarray}
where, due to the ETCR (\ref{dbcr}), the vector coefficients $\Evt_{\om n}(\rv)$ and $\Hvt_{\om n}(\rv)$ are related to $\Dv_{\om n}(\rv)$ and $\Bv_{\om n}(\rv)$ through
\begin{equation}
-i\om\Dv_{\om n}(\rv)=c\;\curl\Hvt_{\om n}(\rv),\qquad -i\om\Bv_{\om n}(\rv)=-c\;\curl\Evt_{\om n}(\rv).
\label{dbehprel}
\end{equation}
We will eventually solve for the coefficients $\Dv_{\om n}(\rv)$ and $\Bv_{\om n}(\rv)$. Thus (\ref{dbehprel}) only defines the transverse parts of $\Evt_{\om n}(\rv)$ and $\Hvt_{\om n}(\rv)$, with the longitudinal parts being unconstrained. Convenient choices for $\Evt_{\om n}(\rv)$ and $\Hvt_{\om n}(\rv)$ are made in (\ref{eptdef}) and (\ref{hptdef}) below so that they correspond to the coefficients of the TP operators in the polariton expansions of $\Eop$ and $\Hop$.

Using the expansions (\ref{doex})--(\ref{sex}), the conditions (\ref{cscr}) and (\ref{sdef}) defining the partitioning of the TP's and LP's may be expressed as requirements upon the expansion coefficients. Substituting (\ref{doex})--(\ref{bex}) in (\ref{hint}) we may reexpress the definition (\ref{sdef}) of the LP's as
\begin{equation}
\intV\;\rhv_{\la n'}^*(\rv,\om')\cdot\left[\Lvb^{\rm e*}_\la(\rv,\om')\cdot\Dv_{\om n}(\rv)+\Lvb^{\rm m*}_\la(\rv,\om')\cdot\Bv_{\om n}(\rv)\right]=0,
\label{rhocond}
\end{equation}
for all $\la$, $n,n'$, and $\om,\om'$. Likewise, by substituting (\ref{cex}) and (\ref{sex}) into the condition (\ref{cscr}) and evaluating the commutator we may reexpress the formal separation of the TP's and the LP's as
\begin{equation}
\intV\;\rhv^*_{\la n'}(\rv,\om')\cdot\av_{\la n}(\rv,\om,\om')=0,
\label{rhoacond}
\end{equation}
for all $\la$, $n,n'$ and $\om,\om'$. Use of (\ref{sex}) in (\ref{sscr}) implies
\begin{equation}
\intV\;\rhv^*_{\la n}(\rv,\om)\cdot\rhv_{\la n'}(\rv,\om)=\delta_{nn'}.
\label{rhvnorm}
\end{equation}

Inserting the expansions (\ref{doex})--(\ref{bex}) into the dynamical equation for the medium operators (\ref{bmdot}) and evaluating the ETCR with $\copa_{\om n}$ and $\cop_{\om n}$ we obtain
\begin{eqnarray}
\left(\om'-\om\right)\av_{\la n}(\rv,\om,\om')&=&\hbsri\left[\Lvb^{\rm e*}_\la(\rv,\om')\cdot\Dv_{\om n}(\rv)+\Lvb^{\rm m*}_\la(\rv,\om')\cdot\Bv_{\om n}(\rv)\right],
\label{aveq}\\
\left(\om'+\om\right)\bv_{\la n}(\rv,\om,\om')&=&\hbsri\left[\Lvb^{\rm e}_\la(\rv,\om')\cdot\Dv_{\om n}(\rv)+\Lvb^{\rm m}_\la(\rv,\om')\cdot\Bv_{\om n}(\rv)\right].
\label{bveq}
\end{eqnarray}
Following the approach of Fano \cite{Fano:1961} as extended by Bhat and Sipe \cite{Bhat:2006} we may use (\ref{rhocond}) and (\ref{rhoacond}) to obtain the solutions to (\ref{aveq}) and (\ref{bveq}) as
\begin{eqnarray}
\av_{\la n}(\rv,\om,\om')&=&\hbsri\left[\mathscr{P}\frac{1}{\om'-\om}+Z_{\om n}\;\delta(\om'-\om)\right]\nonumber\\
&&\hspace{2cm}\times\left[\Lvb^{\rm e*}_\la(\rv,\om')\cdot\Dv_{\om n}(\rv)+\Lvb^{\rm m*}_\la(\rv,\om')\cdot\Bv_{\om n}(\rv)\right],
\label{avsol}\\
\bv_{\la n}(\rv,\om,\om')&=&\hbsri\frac{1}{\om'+\om}\left[\Lveb_\la(\rv,\om')\cdot\Dv_{\om n}(\rv)+\Lvmb_\la(\rv,\om')\cdot\Bv_{\om n}(\rv)\right].
\label{bvsol}
\end{eqnarray}
The complex scalar $Z_{\om n}$ parametrizes the resonant interaction between the electromagnetic field and the medium as a whole and is therefore independent of $\lambda$. It represents a generalized contribution from the pole at $\om=\om'$ originally introduced by Dirac \cite{Dirac:1927}. We find that $Z_{\om n}$, which is undetermined at this point, emerges as the eigenvalue associated with solutions of a generalized Hermitian eigenvalue problem. No pole is considered in obtaining (\ref{bvsol}) since $\om$ and $\om'$ are positive.

Thus, the expressions (\ref{avsol}) and (\ref{bvsol}) have reduced the unknown quantities associated with the expansions (\ref{doex})--(\ref{bex}) to just the vector fields $\rhv_{\la n}(\rv,\om)$, $\Dv_{\om n}(\rv)$, and $\Bv_{\om n}(\rv)$, and the scalar $Z_{\om n}$. The procedure for determining $\rhv_{\la n}(\rv,\om)$ was given by Bhat and Sipe for a dielectric medium \cite{Bhat:2006}. Although we consider a more general coupling here, similar principles apply and solutions for $\rhv_{\la n}(\rv,\om)$ may be obtained from consideration of (\ref{rhocond}) and (\ref{rhvnorm}), along with the transversality of $\Dv_{\om n}(\rv)$ and $\Bv_{\om n}(\rv)$ (which is derived below). However, we do not consider this solution  procedure explicitly here as only the transverse electromagnetic field operators correspond to optically measurable variables, and any coupling between the electromagnetic field and an atomic system, for instance, is mediated by the same. We therefore restrict our attention to the TP operators and the explicit construction of $\Dv_{\om n}(\rv)$ and $\Bv_{\om n}(\rv)$, as these alone contribute to $\Dop$ and $\Bop$ through (\ref{doex}) and (\ref{boex}).

\subsection{\label{tpmfsection}TP mode fields}

The vector fields $\Dv_{\om n}(\rv)$ and $\Bv_{\om n}(\rv)$ are of particular importance as they play the r\^ole of TP mode fields. From (\ref{doex}) and (\ref{boex}) it may be observed that, since the time evolution and commutation properties of the TP operators are known, determination of the TP mode fields constitutes a complete description of the electromagnetic field operators $\Dop$ and $\Bop$.

To determine the TP mode fields we insert (\ref{doex})--(\ref{bex}) into (\ref{ddot}) and (\ref{bdot}) and evaluate the ETCR of both sides with $\copa_{\om n}$ to yield
\begin{eqnarray}
-i\om\;\Dv_{\om n}(\rv)&=&c\;\curl\Bv_{\om n}(\rv)\nonumber\\
&&-c\;\curl\hbsr\sum_\la\inth\dom'\left[\Lvm_\la(\rv,\om')\cdot\av_{\la n}(\rv,\om,\om')\right.\nonumber\\
&&\hspace{5cm}\left.+\Lv^{\rm m*}_\la(\rv,\om')\cdot\bv_{\la n}(\rv,\om,\om')\right],
\label{dpdot}\\
-i\om\;\Bv_{\om n}(\rv)&=&-c\;\curl\Dv_{\om n}(\rv)\nonumber\\
&&+c\;\curl\hbsr\sum_\la\inth\dom'\left[\Lve_\la(\rv,\om')\cdot\av_{\la n}(\rv,\om,\om')\right.\nonumber\\
&&\hspace{5cm}\left.+\Lv^{\rm e*}_\la(\rv,\om')\cdot\bv_{\la n}(\rv,\om,\om')\right].
\label{bpdot}
\end{eqnarray}
Substituting Eqs.~(\ref{avsol}) and (\ref{bvsol}) into Eqs.~(\ref{dpdot}) and (\ref{bpdot}), and recalling the definitions in (\ref{gvparts}), (\ref{ghldef}), and (\ref{galdef}), then leads to the TP mode field equations
\begin{eqnarray}
-i\om\;\Dv_{\om n}(\rv)&=&c\;\curl\left\{\left[1-\Gvh^{\rm mm}(\rv,\om)\right]\cdot\Bv_{\om n}(\rv)-\Gvbh^{\rm em*}(\rv,\om)\cdot\Dv_{\om n}(\rv)\right\}\nonumber\\
&&-c\frac{Z_{\om n}}{i\pi}\curl\left[\Gva^{\rm mm}(\rv,\om)\cdot\Bv_{\om n}(\rv)-\Gvba^{\rm em*}(\rv,\om)\cdot\Dv_{\om n}(\rv)\right],
\label{dpeq}\\
-i\om\;\Bv_{\om n}(\rv)&=&-c\;\curl\left\{\left[1-\Gvh^{\rm ee}(\rv,\om)\right]\cdot\Dv_{\om n}-\Gvh^{\rm em}(\rv,\om)\cdot\Bv_{\om n}\right\}\nonumber\\
&&+c\frac{Z_{\om n}}{i\pi}\curl\left[\Gva^{\rm ee}(\rv,\om)\cdot\Dv_{\om n}(\rv)+\Gva^{\rm em}(\rv,\om)\cdot\Bv_{\om n}(\rv)\right].
\label{bpeq}
\end{eqnarray}
If we set $Z_{\om n}=i\pi$ these equations correspond to the classical, source free Maxwell curl equations in the frequency domain, with the associated complex frequency solutions. However, this is inconsistent with unitary evolution. Instead we seek complex solutions for $Z_{\om n}$ which determine an augmented medium response, thus ensuring the reality of the polariton frequency $\om$. From a related perspective, it is shown below in Section \ref{othersection} that the deviation of $Z_{\om n}$ from a value of $i\pi$ determines the contribution of the corresponding polariton mode to the noise polarization and magnetization.

A consequence of (\ref{dpeq}) and (\ref{bpeq}) are the conditions $\dive\Dv_{\lv\om}(\rv)=\dive\Bv_{\lv\om}(\rv)=0$, from which (\ref{qdiveqs}) follows. Thus, construction of the field operators $\Dop$ and $\Bop$ according to (\ref{doex}) and (\ref{boex}) ensures their transversality, and the condition (\ref{qdiveqs}) need no longer be enforced explicitily as an initial condition. Comparing (\ref{dbehprel}) with (\ref{dpeq}) and (\ref{bpeq}) we are free to identify
\begin{eqnarray}
\Evt_{\lv\om}(\rv)&=&\left[1-\Gvh^{\rm ee}(\rv,\om)\right]\cdot\Dv_{\lv\om}(\rv)-\Gvh^{\rm em}(\rv,\om)\cdot\Bv_{\lv,\om}(\rv)\nonumber\\
&&-c\frac{Z_{\lv\om}}{i\pi}\left[\Gva^{\rm ee}(\rv,\om)\cdot\Dv_{\lv\om}(\rv)+\Gva^{\rm em}(\rv,\om)\cdot\Bv_{\lv\om}(\rv)\right],
\label{eptdef}\\
\Hvt_{\lv\om}(\rv)&=&\left[1-\Gvh^{\rm mm}(\rv,\om)\right]\cdot\Bv_{\lv\om}(\rv)-\Gvh^{\rm me}(\rv,\om)\cdot\Dv_{\lv\om}(\rv)\nonumber\\
&&-c\frac{Z_{\lv\om}}{i\pi}\left[\Gva^{\rm mm}(\rv,\om)\cdot\Bv_{\lv\om}(\rv)+\Gva^{\rm me}(\rv,\om)\cdot\Dv_{\lv\om}(\rv)\right].
\label{hptdef}
\end{eqnarray}
Arbitrary additional terms corresponding to longitudinal fields may be added to the RHS's of (\ref{eptdef}) and (\ref{hptdef}) without effect upon (\ref{cex}) and (\ref{dbehprel}); here such terms are set to zero without loss of generality.

The pair of equations (\ref{dpeq}) and (\ref{bpeq}) may be recast in the form of a single, generalized Hermitian eigenvalue problem for each value of $\om$ in the domain $(0,\infty)$; viz.,
\begin{eqnarray}
&&\curlv\left\{i\left[\begin{array}{cc}
\mathbf{1}-\Gvh^{\rm ee}(\rv,\om) & -\Gvh^{\rm em}(\rv,\om) \\
-\Gvbh^{\rm em*}(\rv,\om) & \mathbf{1}-\Gvh^{\rm mm}(\rv,\om)
\end{array}\right]
\left[\begin{array}{c}
\Dv_{\om n}(\rv) \\ \Bv_{\om n}(\rv)
\end{array}\right]\right\}
+\frac{\om}{c}\left[\begin{array}{c}
\Dv_{\om n}(\rv) \\ \Bv_{\om n}(\rv)
\end{array}\right]\nonumber\\
&=&\frac{Z_{\om n}}{\pi}\curlv\left\{\left[\begin{array}{cc}
\Gva^{\rm ee}(\rv,\om) & \Gva^{\rm em}(\rv,\om) \\
-\Gvba^{\rm em*}(\rv,\om) & \Gva^{\rm mm}(\rv,\om)
\end{array}\right]
\left[\begin{array}{c}
\Dv_{\om n}(\rv) \\ \Bv_{\om n}(\rv)
\end{array}\right]\right\}.
\label{ghep}
\end{eqnarray}
Mode field solutions and their corresponding, generally complex, eigenvalues $Z_{\om n}$ are labeled by the index $n$. To obtain the normalization condition for the solutions to (\ref{ghep}), and thus the TP mode fields, we substitute the expansion (\ref{cex}) into the ETCR (\ref{cccr}), which we then evaluate using the expressions for the expansion coefficients (\ref{avsol}), (\ref{bvsol}), (\ref{eptdef}) and (\ref{hptdef}), with the result
\begin{eqnarray}
&&\left[\cop_{\om n}(t),\copa_{\om'n'}(t)\right]\nonumber\\
&=&-i2\pi\delta(\om-\om')\;\frac{Z^*_{\om n'}Z_{\om n}+\pi^2}{2\hbar\pi^2}\nonumber\\
&&\hspace{1cm}\times\intV\left\{\left[\begin{array}{cc}
\Dv^*_{\om n'}(\rv) & \Bv^*_{\om n'}(\rv)
\end{array}\right]
\left[\begin{array}{cc}
\Gva^{\rm ee}(\rv,\om) & \Gva^{\rm em}(\rv,\om) \\
-\Gvba^{\rm em*}(\rv,\om) & \Gva^{\rm mm}(\rv,\om)
\end{array}\right]
\left[\begin{array}{c}
\Dv_{\om n}(\rv) \\ \Bv_{\om n}(\rv)
\end{array}\right]\right\}.
\label{cccreval}
\end{eqnarray}
In obtaining (\ref{cccreval}) we have exploited the identity \cite{Fano:1961}
\begin{eqnarray}
&&\mathscr{P}\frac{1}{\om'-\om}\mathscr{P}\frac{1}{\om''-\om}+\mathscr{P}\frac{1}{\om-\om'}\mathscr{P}\frac{1}{\om''-\om'}+\mathscr{P}\frac{1}{\om-\om''}\mathscr{P}\frac{1}{\om'-\om''}\nonumber\\&=&\pi^2\delta(\om-\om')\;\delta\left(\frac{1}{2}(\om+\om')-\om''\right).
\label{fanoid}
\end{eqnarray}
Comparison of (\ref{cccreval}) with the RHS of (\ref{cccr}) then implies
\begin{eqnarray}
&&-i\frac{\om}{2}\left(\frac{|Z_{\om n}|^2}{\pi^2}+1\right)\nonumber\\
&&\times\intV\left\{\left[\begin{array}{cc}
\Dv^*_{\om n}(\rv) & \Bv^*_{\om n'}(\rv)
\end{array}\right]
\left[\begin{array}{cc}
\Gva^{\rm ee}(\rv,\om) & \Gva^{\rm em}(\rv,\om) \\
-\Gvba^{\rm em*}(\rv,\om) & \Gva^{\rm mm}(\rv,\om)
\end{array}\right]
\left[\begin{array}{c}
\Dv_{\om n}(\rv) \\ \Bv_{\om n}(\rv)
\end{array}\right]\right\}\nonumber\\
&=&\hbar\om\;\delta_{nn'}.
\label{evnorm}
\end{eqnarray}
The procedure for constructing the $\Dop$ and $\Bop$ operators is then as follows. For each value of $\om>0$ we solve (\ref{ghep}) with appropriate boundary conditions to obtain a discrete set of mode fields $\Dv_{\om n}(\rv)$ and $\Bv_{\om n}(\rv)$, normalized according to (\ref{evnorm}), with eigenvalues $Z_{\om n}$. Combined with the properties of the TP operators, these solutions then provide a complete description of the transverse electromagnetic field operators $\Dop$ and $\Bop$ through (\ref{doex}) and (\ref{boex}). This description relies only upon the macroscopic susceptibility tensors and is therefore free of the ambiguities associated with the coupling tensors $\Lv^{\si}_\la(\rv,\om)$.

\subsection{\label{othersection}Other field operators}
In the modal polariton picture the TP mode fields may be employed to obtain explicit expressions for the noise operators. Specifically, by substituting (\ref{avsol}) and (\ref{bvsol}) into (\ref{bex}), we may use the result to rewrite (\ref{podef}) and (\ref{modef}). Comparison with (\ref{podefg}) and (\ref{modefg}) in the Fourier domain then yields
\begin{eqnarray}
\Pop^{\rm(n)}(\rv,\omega)&=&\sum_n\left(\frac{Z_{\omega n}}{i\pi}-1\right)\left[\Gva^{\rm ee}(\rv,\omega)\cdot\Dv_{\omega n}(\rv)+\Gva^{\rm em}(\rv,\omega)\cdot\Bv_{\omega n}(\rv)\right]\;\cop_{\omega n}\nonumber\\
&&+\hbsr\sum_\la\sum_n\Lve_\la(\rv,\omega)\cdot\rhv_{\la n}(\rv,\omega)\;\sop_{\la\omega n},
\label{pnodefc}\\
\Mop^{\rm(n)}(\rv,\omega)&=&\sum_n\left(\frac{Z_{\omega n}}{i\pi}-1\right)\left[\Gva^{\rm me}(\rv,\omega)\cdot\Dv_{\omega n}(\rv)+\Gva^{\rm mm}(\rv,\omega)\cdot\Bv_{\omega n}(\rv)\right]\;\cop_{\omega n}\nonumber\\
&&+\hbsr\sum_\la\sum_n\Lvm_\la(\rv,\omega)\cdot\rhv_{\la n}(\rv,\omega)\;\sop_{\la\omega n}.
\label{mnodefc}
\end{eqnarray}
As alluded to in Section \ref{tpmfsection}, these expressions clearly demonstrate how the deviation of the eigenvalue $Z_{\om n}$ from $i\pi$ determines the contribution of each TP mode to the noise operators. We note that those parts of the noise polarization and magnetization operators associated with the TP operators only depend upon the macroscopic susceptibility tensors.

The remaining electromagnetic field operators $\Eop$ and $\Hop$ may now be expressed in the form of mode expansions as follows. Substituting (\ref{doex}), (\ref{boex}), and the inverse Fourier transforms of (\ref{pnodefc}) and (\ref{mnodefc}) into (\ref{podefg}) and (\ref{modefg}), we may use (\ref{pomodef}), (\ref{eptdef}), and (\ref{hptdef}) to write
\begin{eqnarray}
\Eop(\rv,t)&=&\sum_n\inth\dom\;\left[\cop_{\om n}(t)\;\Evt_{\om n}(\rv)+\copa_{\om n}(t)\;\Ev^{\rm T*}_{\om n}(\rv)\right]\nonumber\\
&&+\hbsr\sum_{\la}\sum_n\inth\dom\;\left[\Lve_\la(\rv,\om)\cdot\rhv_{\la n}(\rv,\om)\;\sop_{\la\om n}(t)\right.\nonumber\\
&&\hspace{5cm}\left.+\Lv^{\rm e*}_\la(\rv,\om)\cdot\rhv^*_{\la n}(\rv,\om)\;\sopa_{\la\om n}(t)\right],
\label{eoex}\\
\Hop(\rv,t)&=&\sum_n\inth\dom\;\left[\cop_{\om n}(t)\;\Hvt_{\om n}(\rv)+\copa_{\om n}(t)\;\Hv^{\rm T*}_{\om n}(\rv)\right]\nonumber\\
&&+\hbsr\sum_\la\sum_n\inth\dom\;\left[\Lvm_\la(\rv,\om)\cdot\rhv_{\la n}(\rv,\om)\;\sop_{\la\om n}(t)\right.\nonumber\\
&&\hspace{5cm}\left.+\Lv^{\rm m*}_\la(\rv,\om)\cdot\rhv^*_{\la n}(\rv,\om)\;\sopa_{\la\om n}(t)\right].
\label{hoex}
\end{eqnarray}
Thus, in contrast to the transverse field operators $\Dop$ and $\Bop$ which may be completely described by the TP operators and their associated mode fields, the field operators $\Eop$ and $\Hop$ include contributions from the LP operators.

\section{\label{discussionsection}Concluding remarks}
We have presented a canonical quantization of macroscopic electrodynamics in a linear magneto-electric medium. The theory supports a wide class of magneto-electric responses characterized by the macroscopic susceptibility tensors $\Gv^{\si\nu}(\rv,t)$, for $\si,\nu={\rm e,m}$, which are Kramers-Kronig and Onsager consistent. The resultant electromagnetic field operators are expressed in a mode expansion representation and form a natural basis for the study of quantum optics in waveguide and cavity geometries involving dispersive and lossy magneto-electric media, while also paving the way for the inclusion of quantum optical nonlinearities in such structures.

In practice, the available measured or modeled quantities are the susceptibility tensors $\Gv^{\si\nu}$. Thus (\ref{galdef}) may be considered as a definition of the coupling tensors $\Lv^\si_\la$. However, the set of $\Gv^{\si\nu}$'s encompassed by our model is restricted. Specifically, the form of (\ref{galdef}) implies an interdependency between the various susceptibility tensors due to the fact that the magneto-electric response is constructed from the same coupling tensors as the dielectric and magnetic terms. Using the same methods as Horsley \cite{Horsley:2011} it may be shown that for any number of subsystems $N$, the following equality is satisfied:
\begin{equation}
|\Gv^{\rm em}_{{\rm A}ij}(\rv,\omega)|^2\le|\Gv_{{\rm A}ii}^{\rm ee}(\rv,\omega)||\Gv_{{\rm A}jj}^{\rm mm}(\rv,\omega)|.
\label{gaineq}
\end{equation}
Thus, all media to which the present quantum theory applies must exhibit responses that satisfy the condition (\ref{gaineq}).

As an example of the additional freedoms obtained by increasing the number of subsystems with which the medium is modeled, we first consider the case where $N=1$. If the medium is isotropic we have
\begin{eqnarray}
\Gva^{\rm ee}(\rv,\omega)&=&i\mathscr{I}\left\{\Gamma^{\rm ee}(\rv,\omega)\right\}\mathbf{1}=\frac{i}{2}\frac{\omega}{|\omega|}\Lv_1^{\rm e}(\rv,|\omega|)\cdot\Lvb_1^{\rm e*}(\rv,|\omega|),
\label{gaee}\\
\Gva^{\rm mm}(\rv,\omega)&=&i\mathscr{I}\left\{\Gamma^{\rm mm}(\rv,\omega)\right\}\mathbf{1}=\frac{i}{2}\frac{\omega}{|\omega|}\Lv_1^{\rm m}(\rv,|\omega|)\cdot\Lvb_1^{\rm m*}(\rv,|\omega|),
\label{gamm}
\end{eqnarray}
where $\Gamma^{\rm ee}$ and $\Gamma^{\rm mm}$ are scalar functions and $\mathbf{1}$ is the 2nd rank tensor with Cartesian components $\delta_{ij}$. From (\ref{gaee}) and (\ref{gamm}) acceptable forms of the coupling tensors are given to within an arbitrary unitary transformation by
\begin{eqnarray}
\Lve_1(\rv,\om)&=&\left[2\mathscr{I}\left\{\Gamma^{\rm ee}(\rv,\om)\right\}\right]^{\frac{1}{2}}\mathbf{1},
\label{le}\\
\Lvm_1(\rv,\om)&=&\pm i\left[2\mathscr{I}\left\{\Gamma^{\rm mm}(\rv,\om)\right\}\right]^{\frac{1}{2}}\mathbf{1}.
\label{lm}
\end{eqnarray}
The magneto-electric susceptibilities are then fixed by the dielectric and magnetic responses as
\begin{eqnarray}
\Gva^{\rm em}(\rv,\omega)&=&\pm\frac{\omega}{|\omega|}\mathscr{I}\left\{\Gamma^{\rm ee}(\rv,|\omega|)\right\}^{\frac{1}{2}}\mathscr{I}\left\{\Gamma^{\rm mm}(\rv,|\omega|)\right\}^{\frac{1}{2}}\mathbf{1},
\label{gaem}\\
\Gva^{\rm me}(\rv,\omega)&=&-\Gva^{\rm em}(\rv,\omega).
\label{game}
\end{eqnarray}
This condition is relaxed by extending the model to $N=2$ for which we have the general tensor relations
\begin{eqnarray}
\Gva^{\rm ee}(\rv,\omega)&=&\frac{i}{2}\frac{\omega}{|\omega|}\left[\Lv^{\rm e}_1(\rv,|\omega|)\cdot\Lvb_1^{\rm e*}(\rv,|\omega|)+\Lv^{\rm e}_2(\rv,|\omega|)\cdot\Lvb_2^{\rm e*}(\rv,|\omega|)\right],\nonumber\\
\Gva^{\rm mm}(\rv,\omega)&=&\frac{i}{2}\frac{\omega}{|\omega|}\left[\Lv^{\rm m}_1(\rv,|\omega|)\cdot\Lvb_1^{\rm m*}(\rv,|\omega|)+\Lv^{\rm m}_2(\rv,|\omega|)\cdot\Lvb_2^{\rm m*}(\rv,|\omega|)\right],\nonumber\\
\Gva^{\rm em}(\rv,\omega)&=&\frac{i}{2}\frac{\omega}{|\omega|}\left[\Lv^{\rm e}_1(\rv,|\omega|)\cdot\Lvb_1^{\rm m*}(\rv,|\omega|)+\Lv^{\rm e}_2(\rv,|\omega|)\cdot\Lvb_2^{\rm m*}(\rv,|\omega|)\right].
\label{gadef2}
\end{eqnarray}
However, since all four coupling tensors contribute to all the susceptibility tensors in (\ref{gadef2}), properties such as resonances which are present in the electric or magnetic susceptibilities must also manifest in the magneto-electric susceptibility. Now consider $N=3$ in the special case where we set $\Lve_2=\Lvm_1=0$ to maintain the same number of coupling tensors in the model as before. Such a prescription yields
\begin{eqnarray}
\Gva^{\rm ee}(\rv,\omega)&=&\frac{i}{2}\frac{\omega}{|\omega|}\left[\Lv^{\rm e}_1(\rv,|\omega|)\cdot\Lvb_1^{\rm e*}(\rv,|\omega|)+\Lv^{\rm e}_3(\rv,|\omega|)\cdot\Lvb_3^{\rm e*}(\rv,|\omega|)\right],\nonumber\\
\Gva^{\rm mm}(\rv,\omega)&=&\frac{i}{2}\frac{\omega}{|\omega|}\left[\Lv^{\rm m}_2(\rv,|\omega|)\cdot\Lvb_2^{\rm m*}(\rv,|\omega|)+\Lv^{\rm m}_3(\rv,|\omega|)\cdot\Lvb_3^{\rm m*}(\rv,|\omega|)\right],\nonumber\\
\Gva^{\rm em}(\rv,\omega)&=&\frac{i}{2}\frac{\omega}{|\omega|}\Lv^{\rm e}_3(\rv,|\omega|)\cdot\Lvb_3^{\rm m*}(\rv,|\omega|).
\label{gadef3}
\end{eqnarray}
Though the number of coupling tensors are the same as in the $N=2$ case leading to (\ref{gadef2}), in (\ref{gadef3}) there are elements of the electric and magnetic susceptibilities that are constructed from $\Lve_1$ and $\Lvm_2$ and thus decoupled from the magneto-electric susceptibility, which only involves $\Lve_3$ and $\Lvm_3$. This allows for the inclusion of purely electric and magnetic effects which do not manifest in the magneto-electric response.

Finally we recall that in all cases the coupling tensors are defined to within the unitary transformations $\Uv_\la$ introduced in Section \ref{hamiltoniansection}. Inspection of (\ref{galdef}) implies that the ambiguities associated with such transformations, however, have no effect upon the macroscopic susceptibilities which themselves are independent of $\Uv_\la$. The noise operators defined in (\ref{pondeft}) and (\ref{mondeft}), and re-expressed in (\ref{pnodefc}) and (\ref{pnodefc}), are likewise unaffected. More generally, we expect that this independence with respect to $\Uv_\la$ must apply to any physical result of the theory.

In our quantization procedure we have employed the electric induction $\Dop$ and the magnetic induction $\Bop$ as the canonical variables corresponding to the electromagnetic field. This prescription is equivalent to approaches involving the vector potential $\hat{\mathbf{A}}$ and its conjugate momentum $\hat{\boldsymbol{\Pi}}$, with $\Dop=\hat{\boldsymbol{\Pi}}$ and $\Bop=\curl\hat{\mathbf{A}}$ (cf. the ETCR (\ref{dbcr})). In relating the $\Eop$ and $\Hop$ field operators to $\Dop$ and $\Bop$, the classical definitions of the polarization and magnetization are carried over into the quantum domain resulting in (\ref{pomodef}), and the transversality of $\Dop$ and $\Bop$ is preserved. In this respect we are consistent with the work of Suttorp \cite{Suttorp:2007}. In contrast, Philbin \cite{Philbin:2010} only retains the driven part of the polarization and magnetization in (\ref{pomodef}), with the noise operators separated into additional source terms in the divergence equations for $\Dop$ and $\Hop$; the operator $\Dop$ is therefore no longer transverse.

\section*{Acknowledgements}
This research was supported by the Australian Research Council Centre of Excellence for Ultrahigh bandwidth Devices for Optical Systems (project number CE110001018). \mbox{J. E. Sipe} is supported by the National Science and Engineering Research Council of Canada (NSERC).

\bibliography{citations}

\end{document}